\documentclass[12pt]{article}
\usepackage{amsfonts}
\usepackage{enumerate}
\usepackage{graphicx,psfrag,color}
\usepackage{fancyhdr}
\usepackage{amsmath,amsthm,amssymb}
\usepackage{mathrsfs}
\usepackage{indentfirst}
\usepackage{epsfig, epsfig, graphics, lscape, amsbsy, amstext, natbib}
\newtheorem{theorem}{THEOREM}

\newtheorem{lemma}{LEMMA}
\newcommand{\T}{\!\top\!}

\newcommand{\ve}[1]{{\mbox{\boldmath ${#1}$}}}

\begin{document}
\begin{titlepage}

\title{ Fused Mean-variance Filter for Feature Screening
{\small \author{
Xiao-Dong Yan~~and~~Nian-sheng Tang\footnote{Correspondence to:
Dr. Nian-Sheng Tang, Department of Statistics, Yunnan University,
Kunming 650091, P. R. of China. Tel: 86-871-65032416~ Fax: 86-871-65033700~ E-mail: nstang@ynu.edu.cn}   \\
 }}}
\date{}
\maketitle

\thispagestyle{empty}

\noindent {\bf Abstract}: This paper proposes a novel model-free screening procedure for ultrahigh dimensional data analysis.
By utilizing slicing technique which has been successfully applied to continuous variables, we construct a new index called the fused mean-variance for feature screening.
This method has the following merits: (i) it is model-free, i.e., without specifying regression form of predictors and response variable; (ii) it can be used to analyze various types of variables including discrete, categorical and continuous variables; (iii) it still works well even when the covariates/random errors are heavy-tailed or
the predictors are strongly dependent. Under some regularity conditions, we establish the sure screening and rank consistency.
Simulation studies are conducted to assess the performance of the proposed approach.
A real data is used to illustrate the proposed method.
\\

\noindent {\sl  Some key words}: Rank consistency property; Slicing method; Sure screening property; The fused mean-variance; Ultrahigh dimensional data.

\end{titlepage}

\newpage

\section{Introduction}
\renewcommand{\theequation}{1.\arabic{equation}}
\setcounter{equation}{0}
Ultrahigh-dimensional data are often encountered in many research fields such as genomics, bioinformatics, proteomics and high-frequency finance.
In ultrahigh-dimensional data, the number of variables $p$ can grow exponentially with the sample size $n$.
It is recently recognized that only a small number of explanatory variables contribute to the response in the analysis of ultrahigh-dimensional data.
To this end, various model-based feature screening approaches have been proposed to estimate a sparse model and select significant predictors simultaneously for ultrahigh-dimensional data. For example,
Fan $\&$ Lv (2008) proposed a sure independent screening (SIS) and iterated sure independence screening (ISIS) procedure
in the context of linear regression models with Gaussian covariates and responses by ranking the marginal
Pearson correlations; Fan $\&$ Song (2010) extended the SIS procedure to generalized linear models and presented a more general
version of the independent learning with ranking the maximum marginal likelihood estimates or the maximum marginal likelihood itself;
Fan, Feng and Song (2011) developed a nonparametric independence screening (NIS) method with ranking the importance of predictors via the magnitude of nonparametric components
in sparse ultrahigh dimensional additive models;
Chang et al. (2013) proposed a new screening method for linear models and generalized linear models based on the marginal empirical likelihood ratio.
The aforementioned screening methods only work well for the setting that the imposed working models are quite close to the true models (Zhu et al., 2011), but they perform poor in the presence of model misspecification.

To address the aforementioned issue for ultrahigh dimensional data analysis, some model-free feature screening procedures have been developed in recent years.
For example, Zhu et al. (2011) proposed a sure
independent ranking and screening (SIRS) procedure to screen significant predictors under a unified model framework, which includes a wide
variety of commonly used parameter and nonparametric models;
Li, Peng, Zhang and Zhu (2012) proposed a robust rank correlation screening (RRCS) method based on the Kendall $\tau$ correlation
coefficient between response and predictor variables;
Li, Zhong and Zhu (2012) developed a SIS procedure based on the distance correlation;
He et al. (2013) presented a quantile-adaptive-based nonlinear independence screening procedure (QAS); Mai and Zou (2013) proposed
a sure feature screening procedure based on the Kolmogorov distance for binary classification problems, but the Kolmogorov filter screening
is inapplicable when the response variable takes more than two values.
Recently, Cui, Li and Zhong (2015) developed another marginal feature screening procedure for discriminant analysis problem with ultrahigh dimensional
predictors based on empirical conditional distribution function (MVS), which is easily implemented without involving numerical optimization
and is robust to model specification, outliers or heavy tails of the predictors,
but it is studied only for the case that
response variable is categorical and feature is continuous. To overcome the shortcomings of Mai and Zou (2013) and Cui et al. (2015),
Mai $\&$ Zou (2015) proposed a nonparametric model-free screening procedure based on the fused Kolmogorov filter (FKS)
via the slicing technique. The FKS procedure works well for many types of covariates and response variables
such as continuous, discrete and categorical variables, and is invariant under univariate monotone transformation of variable.
But the FKS procedure is computationally intensive in that calculating the Kolmogorov-Smirnov statistic involves numerical optimization.


In this article, our main purpose is to develop an effective and computationally feasible variable screening procedure for ultrahigh dimensional data analysis.
The proposed procedure can be applicable for various types of covariates and response variables, including discrete,
categorical and continuous variables, it is robust to model misspecification, outliers and heavy-tailed data,
and it is easily implemented without involving numerical optimization.
To this end, due to the nature of FKS and MVS procedures, we propose a marginal feature screening procedure via the slicing technique,
which is referred to as the fused mean-variance (FMV) screening,
based on empirical conditional distribution function. We then discuss its asymptotic properties and show the sure screening and rank
consistency properties under general regularity conditions.
The FMV screening has the following merits: (i) it combines the characteristics of the MVS and FKS procedures;
 (ii) it is model-free, i.e., without specifying any regression form of predictor and response variables;
(iii) it has the sure screening property even when predictors are strongly dependent on each other;
(iv) it performs well in the presence of model misspecification, outliers and heavy-tailed data.

The rest of this article is organized as follows. In Section 2, we introduce the FMV method for feature screening.
In Section 3, we study its theoretical properties under some regularity conditions.
In Section 4, we conduct Monte Carlo simulation studies to investigate the finite sample performance of the proposed method. In Section 5,
a real data example is used to illustrate the proposed screening procedure. Technical details are presented in the Appendix.

\section{Method}
\renewcommand{\theequation}{2.\arabic{equation}}
\setcounter{equation}{0}
\subsection{Motivation}

Let $Y$ be a categorical response with $R$ classes $\{y_1,\ldots,y_R\}$, and $X$ be a continuous covariate with a support $\mathbb{R}_{X}$.
Define $F(x)=\mathbb{P}(X\leq x)$ as the unconditional distribution function of $X$, and $F_r(x)=\mathbb{P}(X\leq x|Y=y_r)$ as the conditional distribution function of $X$ given $Y=y_r$.
A variable $X$ is independent of the response variable $Y$ if and only if $F_r(x)=F(x)$ for any $x\in\mathbb{R}_X$ and $r=1,\ldots,R$.
Due to the aforemention fact, Cui, Li and Zhong (2015) considered using the index
$${\rm MV}(X|Y)=E_X[{\rm var}_Y\{F(X|Y)\}]$$
to measure the dependence between $X$ and $Y$, where $F(x|Y)=\mathbb{P}(X\leq x|Y)$.
Cui, Li and Zhong (2015) showed that (i) MV$(X|Y)=\sum_{r=1}^Rp_r\int\{F_r(x)-F(x)\}^2dF(x)$, and (ii) MV$(X|Y)=0$ if and only if $X$ and $Y$ are statistically independent,
where $p_r=\mathbb{P}(Y=y_r)>0$ for $r=1,\ldots,R$. Given the observed data set $\{(X_i,Y_i): i=1,\ldots,n\}$ from the population, an empirical estimator of MV$(X|Y)$ is given by
$$\widehat{\rm MV}(X|Y)=\frac{1}{n}\sum_{r=1}^R\sum_{j=1}^n\hat{p}_r\{\hat{F}_r(X_j)-\hat{F}(X_j)\}^2,$$
where $\hat{p}_r=\frac{1}{n}\sum_{i=1}^nI(Y_i=y_r)$ with $I(\cdot)$ being the indicator function, $\hat{F}(x)=\frac{1}{n}I(X_i\leq x)$, and $\hat{F}_r(x)=\frac{1}{n}\sum_{i=1}^nI(X_i\leq x,Y_i=y_r)/\hat{p}_r$. Cui, Li and Zhong (2015) established the corresponding asymptotic properties for the proposed screening procedure under some regularity conditions,
and demonstrated its satisfactory performance.

Motivated by the success of Cui, Li and Zhong (2015), we want to extend the work of Cui, Li and Zhong (2015) to a continuous response variable
or a general categorical response variable $Y$ taken countable values like Poisson random variable with the support $\mathbb{R}_Y$.
To this end, we consider using the following index
\begin{equation}{\label{MV}}
{\rm MV}_j=E_{X_j}[{\rm var}_Y\{F(X_j|Y)\}]=\iint\{F_j(x|Y=y)-F_j(x)\}^2dF_{j}(x)dF_Y(y)
 \end{equation}
to measure the dependence between $X_j$ and $Y$, where $F_Y(y)=\mathbb{P}(Y\leq y)$, and $F_j(x)=\mathbb{P}(X_j\leq x)$ and $F_j(x|Y=y)$ represents the conditional distribution function of
$X_j$ given $Y$ evaluated at $Y=y$. It is easily shown that MV$_j=0$ if and only if $X_j$ is independent of $Y$, which implies that we can use the MV$_j$ as
a marginal utility for feature screening to characterize both linear and nonlinear relationships in ultrahigh dimensional data analysis.

It is difficult to compute MV$_j$ when $F_j(x)$ or $F_Y(y)$ are unknown. Following the widely used method, we use its empirical version to estimate MV$_j$.
When $Y$ is a categorical response having a growing number of classes in the order of $O(n^{\kappa})$ with some $\kappa>0$,
we can employ the aforementioned screening procedure of Cui et al. (2015) to estimate MV$_j$.
However, it is quite difficult to estimate MV$_j$ when $Y$ is a continuous random variable or a discrete random variable having countable values in that
it involves evaluating $F_j(x|y)$ for all possible values $y$. To address the issue, the widely adopted approach is to approximate MV$_j$ by slicing the response (Mai and Zou, 2015).
To this end, we define the following partition of the support $\mathbb{R}_Y$:
\begin{equation}{\label{slice}}
\mathbb{S}=\begin{Bmatrix}
[a_g,a_{g+1}):a_g<a_{g+1},g=1,\ldots,S
\end{Bmatrix},
 \end{equation}
where $a_1=\inf\{y,F_Y(y)<1\}$ and $a_{S+1}=\sup\{y,F_Y(y)<1\}$. The generalization $[a_g,a_{g+1})$ is called as a slice. We also define a random variable $G=\{1,\ldots,S\}$ such that $G=g$ if and only if $Y$ is in the $g$th slice $[a_g,a_{g+1})$ for $g=1,\ldots,S$.
Particularly, when $Y$ is a discrete variable such as a multiclass variable, i.e., $Y=1,\ldots,S$, we thus take $G=Y$.

Although we can not evaluate $F_j(x|Y=y)$ for all possible values $y$, we can approximate $F_j(x|Y=y)$ on a slice $G=g$ (i.e., $a_g\le Y <a_{g+1}$) by using $F_j^\mathbb{S}(x|G=g)$, where $F_j^\mathbb{S}(x|G=g)=\mathbb{P}(X_j\leq x|G=g)$.
Thus, the sliced MV$_j$ can be approximated by
\begin{equation}{\label{slicemv}}
{\rm MV}_j^\mathbb{S}=\sum\limits_{g=1}^Sp_g^\mathbb{S}\int\{F_j^\mathbb{S}(x|G=g)-F_j(x)\}^2dF_{j}(x),
 \end{equation}
where $p_g^\mathbb{S}=\mathbb{P}(G=g)$ and $F_j^\mathbb{S}(x|G=g)=\mathbb{P}(X_j\le x,G=g)/p_g^\mathbb{S}$.

From (\ref{MV}) and (\ref{slicemv}), it is easily seen that the integral problem of continuous variable is transformed into the tractable sum of discrete variables via the slicing  technique. By Equation (\ref{slicemv}), MV$_j^{\mathbb{S}}$ can be regarded as the weighted average of Cram$\acute{\rm e}$r-von Mises distance between the conditional distribution of $X_j$ given the slice $G=g$ and the unconditional distribution function of $X_j$. When $Y$ is multiclass, the slicing decomposes multiclass problem into pairwise binary problems.
Mai $\&$ Zou (2015) have argued the availability of the slicing technique when $Y$ is a categorical response variable that takes infinite values such as a Poisson random variable.
When $Y$ is continuous, slicing has been become a popular tool for reducing dimension (Li, 1991; Cook $\&$ Weisberg, 1991).

Note that MV$_j^{\mathbb{S}}$ enjoys the same property as MV$_j$ in that MV$_j^{\mathbb{S}}$=0 if and only if $X_j$ is independent of $Y$ when $Y$ takes finite values and each possible value of $Y$ forms a slice. But when $Y$ is continuous, it is a challenging task for demonstrating the equivalence of MV$_j^{\mathbb{S}}$ and MV$_j$ in feature screening.
Thus, the following lemma illustrates the dependence between $Y$ and $X_j$ when $Y$ is continuous.

\begin{lemma}

 (i) The necessary and sufficient condition of independence between $X_j$ and Y is {\rm FMV}$_j^\mathbb{S}=0$ for all possible choices of $\mathbb{S}$.

 (ii) Suppose that $X_j$ is not independent of Y and $\mathbb{P}(Y\leq y|X_j=x)$ is not a constant in $x$ for any fixed $y\in\mathbb{R}_Y$,
 we have ${\rm MV}_j^\mathbb{S}\ne0$ for any $\mathbb{S}$.

 (iii) Assume that $F_j(x|y)$ is continuous in $y$. If $\max_{g=1,\cdots,S}\mathbb{P}(G=g)\rightarrow 0$ and $\lim_{S\rightarrow \infty} S\mathbb{P}(G=g)\rightarrow 1$, thus we have ${\rm MV}_j^\mathbb{S}\rightarrow {\rm MV}_j$. Therefore, for $X_j\in \mathbb{R}_{X_j}$ which is not independent of $Y$,
 ${\rm MV}_j^\mathbb{S}>0$ for sufficiently large $S$.
\end{lemma}

Lemma 1 (i) and (ii) theoretically show that ${\rm MV}_j^{\mathbb{S}}$ can be used to measure the correlation between $X_j$ and $Y$, which indicates that
${\rm MV}_j^{\mathbb{S}}$ can be regarded as a surrogate of ${\rm MV}_j$ for variable screening in ultrahigh dimensional data analysis. Moreover, Lemma 1 (iii) demonstrates that ${\rm MV}_j^{\mathbb{S}}$ could be a better measure of dependence between $X_j$ and $Y$ for feature screening than ${\rm MV}_j$.

\subsection{Estimation procedure}

In this section, we focus on the sample version of MV$_j^{\mathbb{S}}$.
Let $\{(X_{ij},Y_i): 1\le i\le n\}$ be a random sample of size $n$ from the population $(X_j,Y)$.
Define $\widehat{p}_g^\mathbb{S}=\frac{1}{n}\sum_{k=1}^n I(a_g\le Y_k <a_{g+1})$, $\widehat{F}_j(x)=\frac{1}{n}\sum_{k=1}^n I(X_{kj}\le x)$ and
$\widehat{F}_j^\mathbb{S}(x|G=g)=\frac{1}{n} \sum_{k=1}^n I(X_{kj}\le x, a_g\le Y_k<a_{g+1})/\widehat{p}_g^\mathbb{S}$.  Following the widely adopted method (e.g., Cui et al., 2015), given a partition $\mathbb{S}$ of the support $\mathbb{R}_Y$ of $Y$,
${\rm MV}_j^{\mathbb{S}}$ can be estimated by its sample version:
\begin{equation}{\label{FMVet}}
\widehat{\rm MV}_j^{\mathbb{S}}=\frac{1}{n}\sum_{i=1}^n\sum_{g=1}^S \widehat{p}_g^\mathbb{S}\{\widehat{F}_j^\mathbb{S}(X_{ij}|G=g)-\widehat{F}_j(X_{ij})\}^2.
\end{equation}

When $Y$ is a multilevel categorical response, $\widehat{\rm MV}_j^{\mathbb{S}}$ defined in Equation (\ref{FMVet}) is just the sample counterpart of the screening index
defined in Equation (2.1) of Cui et al. (2015), which indicates that we extend Cui et al.'s (2015) method to the case that $Y$ is continuous.
When $Y$ is discrete and takes infinite possible values, we take $G=Y+1$ if $Y<S-1$ and $G=S$ if $Y\geq S-1$ (Mai and Zou, 2015),
which indicates that we can still use the above defined $\widehat{\rm MV}_j^{\mathbb{S}}$ to approximate MV$_j$.
However, for a continuous response $Y$, a nature question is how to determine the number (i.e., $S$) of slices and partition the support $\mathbb{R}_Y$ of $Y$ into $S$ slices in applications.
Many authors have discussed the issue in sufficient dimension reduction literature. For example, see
Li (1991), Hsing $\&$ Caroll (1992), Zhu  $\&$ Ng (1995), and Mai and Zou (2015). Although several authors pointed out that the selection of the number of slices has little effect on variable screening results, significant improvement can be obtained by fusion (Cook and Zhang, 2014; Mai and Zou, 2015). To this end,
we consider $K$ different slice schemes, and compute ${\rm MV}_j^{\mathbb{S}}$ for each of $K$ slice schemes and then take the sum of ${\rm MV}_j^{\mathbb{S}_1},\ldots,{\rm MV}_j^{\mathbb{S}_K}$. Therefore, we propose the fused mean-variance filter given by
 \begin{eqnarray}\label{fusedmvest}
  \widehat{{\rm FMV}}_j=\sum_{k=1}^K\widehat{{\rm MV}}_j^{\mathbb{S}_k}
 \end{eqnarray}
 as an estimate of
\begin{eqnarray}\label{fusedmv}
   {\rm FMV}_j=\sum\limits_{k=1}^K{\rm MV}_j^{\mathbb{S}_k},
 \end{eqnarray}
 where $\mathbb{S}_k$ represents the $k$th slice scheme containing $S_k$ intervals for $k=1,\ldots,K$. Generally, $S_k$ is selected such that $S_k\leq \lceil\log(n)\rceil$ for all $k\in\{1,\ldots,K\}$ (Mai and Zou, 2015), where $\lceil a\rceil$ represents the integer part of real number $a$.

 To evaluate  $\widehat{{\rm FMV}}_j$ in (\ref{fusedmvest}), we still need to determine ${\mathbb{S}_k}$. If the distribution of $Y$ is known, then we can consider an oracle uniform slicing to form partitions ${\mathbb{S}_k}$ through $S_k$ intervals with $a_g^k=F_Y^{-1}((g-1)/S_k)$, $k=1, \ldots, S_k$. In practice, $F_Y(y)$ is unknown and can be estimated by the empirical distribution estimator $\widehat{F}_Y(y)$. So, we can estimate $a_g^k$ by
$\widehat{a}_g^k=\widehat{F}_Y^{-1}((g-1)/S_k)$. Write
$$\widehat{\mathbb{S}}_k=\{[\hat a_g^k, \hat a_{g+1}^k): \; \hat a_g^k< \hat a_{g+1}^k, g=1,\ldots, S_k\}$$
as an intuitive  uniform slicing. For the oracle uniform slicing, set
\begin{eqnarray}\label{ora1}
\omega_j^{\circ}=\sum_{k=1}^{K}{MV}_j^{\mathbb{S}_k} \quad \mbox{and} \quad \widehat \omega_j^{\circ}=\sum_{k=1}^{K}\widehat{MV}_j^{\mathbb{S}_k}
 \end{eqnarray}
and  for the intuitive uniform slicing, set
\begin{eqnarray}\label{ora2}
\omega_j=\sum_{k=1}^{K}{MV}_j^{\widehat{\mathbb{S}}_k} \quad \mbox{and} \quad \widehat \omega_j=\sum_{k=1}^{K}\widehat{MV}_j^{\widehat{\mathbb{S}}_k}.
\end{eqnarray}



\section{Theoretical properties}

In this section, we establish the sure screening and rank consistency properties of the proposed fused mean-variance feature screening procedure.

Without specifying a regression model of response variable $Y$ and covariates $X=\{X_1,\ldots,X_p\}$, we define the active predictor subset as
\begin{equation}\label{Dd}
  \mathcal{D}=\{j:F(y|X)\;{\rm functionally\;depends\;on}\;{\sl X_j}\;{\rm for\;some}\;{\sl y}\},
\end{equation}
and use $\mathbb{I}=\{1,2,\ldots,p\}\backslash \mathcal{D}$ to represent the inactive predictor subset, where $p\gg n$ and $n$ is the sample size.

In ultrahigh dimensional data analysis, the sparsity assumption is $p\gg|\mathcal{D}|$.
Hence, our main goal is to find a reduced model with an appropriate scale which can almost fully
contain $\mathcal{D}$ via an independence screening method. To this end, we use the above defined feature screening index MV$_j$
to screening important predictors among $X_j$'s for $j=1,\ldots,p$. It follows from the preceding argument on MV$_j$ that MV$_j=0$
if and only if $X_j$ is independent of $Y$ (i.e., $X_j$ is not an important predictor for fitting $Y$).
Thus, when $\{X_j: j\in\mathcal{D}\}$ is independent of $\{X_j: j\in \mathbb{I}\}$,
MV$_j$ can be regarded as an effective measure for discriminating the active and inactive predictor subsets in that MV$_j>0$
for $j\in \mathcal{D}$ and MV$_j=0$ for $j\in\mathbb{I}$. Clearly, the proposed screening procedure is model-free and
can be adopted to analyze the linear and nonlinear relationships between the response variable and predictors.

In what follows, we introduce  marginal filter
 $\widehat{\omega}_j^{\circ}$ in (\ref{ora1}) to screen the active predictors under the  oracle uniform slicing. The screening subset for this slice pattern is defined as
$$\widehat{\mathcal{D}}_{oracle}=\{j:\widehat{\omega}_j^{\circ}\;{\rm is\;among\;the\;} d_nth\;{\rm largest}\}$$
 for a given size $d_n<n$, where $d_n$ is the predefined positive integer.
 When we choose $\widehat{\omega}_j$ in (\ref{ora2}) as a marginal utility to measure the importance of $X_j$ for response variable $Y$,
 $\mathcal{D}$ can be estimated by
\begin{equation}\label{Ddd}
  \widehat{\mathcal{D}}=\{j: \widehat{\omega}_j \;{\rm is\; among\; the\;} d_nth\;{\rm largest}\}.
\end{equation}
We call the above defined screening procedure (\ref{Ddd}) as the fused MV-based sure independent screening (FMV-SIS).

Now we study the asymptotic properties of the proposed FMV-SIS. To show the sure screening property of the FMV-SIS, we consider the following regularity conditions:


\begin{itemize}
\item [(C1)] There exists a set $\mathcal{E}$ such that $\mathcal{D}\subset \mathcal{E}$ and $\Delta_\mathcal{E}=\min_k\{\min\limits_{j\in\mathcal{E}}{\rm MV}_j^{\mathbb{S}_k}-\max\limits_{j\notin\mathcal{E}}{\rm MV}_j^{\mathbb{S}_k}\}>0;$

\item [(C2)] Let $S_{\min}=\min_k\{S_k\}$. For any $b_1$ and $b_2$ such that $\mathbb{P}(Y\in [b_1,b_2))\leq (1+\Delta_\mathcal{E})/S_{\min}$, we have $\sup_{x\in\mathbb{R}_{X_j}}|F_j(x|y_1)-F_j(x|y_2)|\leq \Delta_\mathcal{E}/8$ for any $j\in\{1,\ldots,p\}$ and $y_1, y_2\in[b_1,b_2)$.
    Assume that $S_k=O(n^\kappa)$ for $\kappa\ge 0$.
\end{itemize}

The above presented regularity conditions are weaker than those for the SIS (Fan $\&$ Lv, 2008)
because (i) we do not require specifying the linear regression function of $Y$ on $X$;
(ii) in comparison to DCS (Li et al., 2012), we make no assumptions on the moments of predictors.
Therefore, the FMV-SIS is expected to be robust to heavy tailed distribution of
predictors and outliers. Moreover, without assuming any form of the dependence of $Y$ on $X$, thus
the FMV-SIS will be more flexible than the NIS and QAS.
Condition (C1) is similar to that given in Mai and Zou (2015), which is used to guarantee that jointly important predictors should also be marginally important.
Condition (C2) is similar to that given in Mai and Zou (2015), which is slightly stronger than assuming $F_j(x|Y=y)$ to be continuous in $y$, and is assumed
to guarantee that $\omega_j^{\circ}$ can be consistently approximated by $\widehat\omega_j$ for $j\in\{1,\ldots,p\}$. Assumption $S_k=O(n^\kappa)$ given in Condition (C2) allows the diverging number of the slices of the
response with the sample size $n$.

\begin{theorem}
{\rm (i)} Assume Condition (C2) hold,  
\begin{equation}{\label{theorem22}}
\mathbb{P}(\max\limits_{1\leq j\leq p}|\widehat{\omega}_j-\omega_j^\circ|\leq K\Delta_{\mathcal{E}})\ge CKp\exp\{-Cn^{1-\kappa}\Delta_{\mathcal{E}}^2+(1+\kappa)\log(n)\}.
\end{equation}
{\rm (ii)}(Sure Screening Property)

Under Condition (C1), 
\begin{equation}{\label{theorem2i}}
\mathbb{P}(\mathcal{D}\subset\widehat{\mathcal{D}}_{(oracle)})\ge 1-CKp\exp\{-Cn^{1-\kappa}\Delta_{\mathcal{E}}^2+(1+\kappa)\log(n)\},
\end{equation}
Under Conditions (C1) and (C2), we have
\begin{equation}{\label{theorem2ii}}
\mathbb{P}(\mathcal{D}\subset\widehat{\mathcal{D}})\ge 1-CKp\exp\{-Cn^{1-\kappa}\Delta_{\mathcal{E}}^2+(1+\kappa)\log(n)\}.
\end{equation}
\end{theorem}
\begin{theorem}(Ranking Consistency Property)
Assume  Conditions (C1) and (C2) hold,
$n^{2\kappa-1}=o(\Delta^2_\mathcal{E})$ and $n^{\kappa-1}\log(Kp)=o(\Delta^2_\mathcal{E})$, then $\liminf\limits_{n\rightarrow\infty}\{\min\limits_{j\in\mathcal{D}}\widehat{\omega}_j-\max\limits_{j\not\in\mathcal{D}}\widehat{\omega}_j\}>0.$
\end{theorem}
Under (\ref{theorem22}), if we pre-determine a threshold value $\tau$, and set $\Delta_\mathcal{E}\geq Cn^{-\tau}$ with some constant $C$, FMV-SIS can handle the NP-dimensionality $\log p=O(n^\xi)$, where $\xi <1-2\tau-\kappa$ with $0\leq \tau \leq 1/2$ and $0 \leq \kappa < 1-2\tau$, which depends on the minimum true signal strengthen and the number of slice. If the slice number is not growing, meaning $\kappa=0$, we have
\begin{eqnarray}
  \mathbb{P}\{\underset{1\leq j\leq p}{\max}|\widehat{\omega}_j-\omega_j^\circ|>CKn^{-\tau}\}\leq CK\exp\{-Cn^{1-2\tau}+\log n\}
\end{eqnarray}
for some constant $C>0$. In this case, we can handle the even larger NP-dimensionality $\log p=O(n^{\xi})$, where $\xi<1-2\tau$ with $0\leq\tau<\frac{1}{2}$.
Theorem 2 demonstrates that the  values of $\widehat{\omega}_j$ of active predictors are ranked ahead that of inactive ones with high probability. So  we can separate the active and inactive predictors through taking an ideal thresholding value.

\section{Simulations}
\subsection{Simulation designs}

Several simulation studies are conducted to investigate the performance of the proposed {\rm FMV} method in terms of the following two criteria:
(i) the median of the minimum model sizes (MMSs, i.e., the smallest number of the selected covariates including all the active predictors)
for $300$ repetitions, (ii) standard error (SE) of 300 MMSs.
To implement the proposed FMV method, we take the maximum and minimum numbers of slices to be $\lceil n^{1/3}\rceil$ and $3$ (Cook \& Zhang, 2014), respectively,
where $\lceil a \rceil$ represents the smallest integer being greater than or equal to $a$. For example, for $n=200$, we consider four slice schemes (i.e., $K=4$) and
select $S_k\in \{3,4,5,6\}$ for $k=1,\ldots,4$. When the response is continuous, we slice $Y$ using the $g/S_k$-th sample quantile of $Y_i$'s for $g=1, \cdots, S_k-1$.

For comparison, we also consider the existing eight screening methods including marginal correlation screening (SIS) (Fan $\&$ lv, 2008),
nonparametric independence screening (NIS) (Fan et al., 2011), sure independent ranking and screening (SIRS) (Zhu et al., 2011),
distance correlation screening (DCS) (Li et al., 2012), rank correlation screening (RCS) (Li et al., 2012),
empirical likelihood screening (ELS)(Chang et al., 2013), quantile-adaptive screening (QA)
(He et al., 2013) and the fused Kolmogorov filter screening (FKS) (Mai $\&$ Zou, 2015).

{\bf Experiment 1} (Linear model with $n=200$ and $p=3000$). In this experiment, we consider the following linear regression model:
$Y_i=\ve X_i^{\top}\ve\beta+\epsilon_i$, where $\ve X_i\sim \mathcal{N}_p(\ve 0,\ve\Sigma)$ and it is assumed that $\epsilon_i$ is independent of $\ve X_i$.
The data set $\{(Y_i,\ve X_i): i=1,\ldots,n\}$ is generated from the above considered linear regression model
with the following specifications of $\ve\beta$, $\ve\Sigma$ and $\epsilon_i$.

Case (1a).~ The true values of parameters $\ve\beta$ and $\ve\Sigma$ are respectively taken to be $\ve\beta=(\ve 1_8,\ve 0_{p-8})$ and $\ve\Sigma=(\sigma_{kj})_{3000\times 3000}$ with $\sigma_{kk}=1.0$ and $\sigma_{kj}=0.8^{|k-j|}$ for $k\neq j$;
and it is assumed that $\epsilon_i\sim \mathcal{N}(0,1)$.

Case (1b).~ We consider the same parameter setting as Case (1a) except for $\epsilon_i\sim t_1$, where $t_1$ denotes the $t$-distribution with one degree of freedom.

Case (1c).~ The true values of parameters $\ve\beta$ and $\ve\Sigma$ are respectively taken to $\ve\beta=(2.0,-2.0,\ve 0_{p-2})$ and $\ve\Sigma=(\sigma_{kj})_{3000\times3000}$ with $\sigma_{kk}=0.8$ and $\sigma_{kj}=0$ for $k\neq j$; and it is assumed that $\epsilon_i\sim \mathcal{N}(0,1)$.

Case (1d).~ We take the same parameter setting as Case (1c) except that outliers are created by multiplying $y_{10}$, $y_{30}$, $y_{50}$ and $y_{70}$ by $100$,
and $y_{20}$, $y_{40}$, $y_{60}$ and $y_{80}$ by $-100$. 
\vspace{2mm}

{\bf Experiment 2} (Variable-transformation linear normal model with $n=200$ and $p=3000$, Mai $\&$ Zou (2015)). In this experiment, we consider the following
transformation linear model: $T_y(Y)=\ve{T}(\ve{X})^{\top}\ve\beta+\epsilon$, where $\ve{T}(\ve X)=(T_1(X_1),\ldots,T_p(X_p))^{\top}$, and $T_y(\cdot)$
and $T_1(\cdot),\ldots,T_p(\cdot)$ are strictly monotone univariate transformation functions. It is assumed that $\ve{X}$ follows the multivariate
normal distribution  $\mathcal{N}_p(\ve 0,\ve\Sigma)$, and $\epsilon\sim N(0,\sigma^2)$ is independent of $\ve X$, where $\ve\Sigma=(\sigma_{jk})$ and $\sigma^2=1.0$.
Denote ${\rm CS}(0.8)=(\sigma_{jk})_{p\times p}$ with $\sigma_{jj}=0.8$
and $\sigma_{jk}=0.0$ for $j\not=k\in \{1,\ldots,p\}$, and ${\rm AR}(0.8)=(\sigma_{jk})_{p\times p}$ with $\sigma_{jj}=1.0$ and
$\sigma_{jk}=0.8^{|j-k|}$ for $j\not=k\in \{1,\ldots,p\}$. The data set $\{(Y_i,\ve X_i): i=1,\ldots,n\}$ is generated from the above given transformation
linear model together with the following specifications of $T_y(Y)$, $T_j(X_j)$, $\ve\beta$ and $\ve\Sigma$:

Case (2a).~ $T_y(Y)=Y$, $T_j(X_j)=X_j^{1/9}$, $\ve\beta=(3.0,-3.0,\ve 0_{p-2})$ and $\ve\Sigma={\rm CS}(0.8)$;

Case (2b).~ $T_y(Y)=Y^{1/9}$, $T_j(X_j)=X_j$, $\ve\beta=(3.0,-3.0,\ve 0_{p-2})$ and $\ve\Sigma={\rm CS}(0.8)$;

Case (2c).~ $T_y(Y)=\log(Y)$, $T_j(X_j)=X_j$, $\ve\beta=(\ve 1_8,\ve 0_{p-8})$ and $\ve\Sigma={\rm AR}(0.8)$.
\vspace{2mm}

{\bf Experiment 3} (Single index regression model with $n=200$ and $p=3000$).
In this experiment, the data set $\{(Y_i,\ve X_i): i=1,\ldots,n\}$ is generated from the following single index regression model: $Y_i=(3X_{1i}+2X_{2i}+X_{3i})^3+\epsilon_i$,
where $\ve X_i=(X_{1i},X_{2i},X_{3i})$,
$X_{ji}$'s are independently drawn from the Cauchy distribution for $j=1,2,3$,
and $\epsilon_i$'s are independently sampled from the standard normal distribution $\mathcal{N}(0,1)$.
\vspace{2mm}

{\bf Experiment 4} (Additive regression model with $n=200$ and $p=3000$). In this experiment, the data set $\{(Y_i,\ve X_i): i=1,\ldots,n\}$ is generated from the following model:
$Y_i=4X_{1i}+2\tan(\pi X_{2i}/2)+5X_{3i}^2+\epsilon_i$, where $X_{ji}$'s
are independently generated from the uniform distribution ${\rm Unif}(0,1)$ and $\epsilon_i$'s are independently sampled from the standard normal distribution
$\mathcal{N}(0,1)$, where $\ve X_i=(X_{1i},X_{2i},X_{3i})$.
\vspace{2mm}

{\bf Experiment 5} (Heteroskedastic regression model with $n=200$ and $p=3000$). In this experiment, the data set $\{(Y_i,\ve X_i): i=1,\ldots,n\}$ is generated from the following model: $Y_i=2(X_{1i}+0.8X_{2i}+0.6X_{3i}+0.4X_{4i}+0.2X_{5i})+\exp(X_{20,i}+X_{21,i}+X_{22,i})\epsilon_i$,
where $\epsilon_i$'s are generated from the standard normal distribution $\mathcal{N}(0,1)$, and
$\ve X_i=(X_{1i},\ldots,X_{pi})^{\top}$'s are drawn from the normal distribution $\mathcal{N}_p(\ve 0,\ve\Sigma)$ with $\ve\Sigma=AR(0.8)$.
This model has even been used by Zhu et al. (2014) and He, Wang and Hong(2013) for investigating the performance of the QA screening method.
In their empirical studies, the QA screening method can only detect $5$ active variables for the quantile $\alpha=0.5$, but it can detect $8$ active variables
for other quantiles such as $\alpha=0.75$.
For comparison, we also present the minimum number of predictors that can include all eight active variables for the QA screening method with $\alpha=0.5$.
\vspace{2mm}

{\bf Experiment 6} (Poisson regression model with $n=200$ and $p=3000$).
In this experiment, the data set $\{(Y_i,\ve X_i): i=1,\ldots,n\}$ is generated from the Poisson distribution
$Y_i\sim {\rm Poisson}(\mu_i)$ with $\mu_i=\exp(\ve X_i^{\top}\ve\beta)$, where $\ve\beta=(0.8,-0.8,\ve 0_{p-2})^{\top}$ and
$\ve X_i=(X_{1i},\ldots,X_{pi})^{\top}$ in which
$X_{ji}$'s are independently drawn from the $t$-distribution with one degree of freedom.
For comparison, we calculate the marginal maximum likelihood estimator (MMLE) (Fan $\&$ Song, 2010) for the SIS method.
It is worthwhile to note that the above drawn predictors may be heavy-tailed, and thus $Y_i$'s may have some extremely outliers.
Therefore, for computationally feasible, we discard the observations $(Y_i,\ve X_i)$ with $Y_i>1000$ in evaluating MMLE.
We also calculate the corresponding results for the FMV, fused Kolmogorov filter and DCS methods.
For the FMV and fused Kolmogorov filter methods, we take $G=Y$ if $Y<2$ and $G=3$ if $Y\geq 2$.
\vspace{2mm}

{\bf Experiment 7} (Censored data with $n=400$ and $p=1000$). In this experiment,
the data set $\{(Y_i,\ve X_i): i=1,\ldots,n\}$ is generated from the model:
$Y_i=5g_1(X_{1i})+3g_2(X_{2i})+4g_3(X_{3i})+6g_4(X_{4i})+\sqrt{1.74}\epsilon_i$, which has even been considered by He et al. (2013)
for illustrating their proposed screening method,  where
$\ve X_i=(X_{1i},\ldots,X_{pi})^{\T}$'s are independently generated from the multivariate normal distribution with mean $\ve 0$
and covariance matrix $\ve\Sigma={\rm CS}(1.0)$,
$\epsilon_i$'s are independently sampled from the standard normal distribution $\mathcal{N}(0,1)$,
$g_1(x)=x$, $g_2(x)=(2x-1)^2$, $g_3(x)=\sin(2\pi x)/(2-\sin(2\pi x))$, and $g_4(x)= 0.1\sin(2\pi x)+0.2\cos(2\pi x)+0.3\sin(2\pi x)^2 +0.4\cos(2\pi x)^3 +0.5\sin(2\pi x)^3$.
Let $Y_i^*=\min(Y_i ,C_i )$, where the censoring time $C_i$'s are generated
from the following mixture normal distribution: $0.4\mathcal{N}(-5,4) + 0.1\mathcal{N}(5,1)+0.5\mathcal{N}(55,1)$.
The average censoring proportion is roughly 30\%. Based on the above generated data set $\{(Y_i^*,\ve X_i): i=1,\ldots,n\}$, we compute the results for the above proposed FMV method.
For comparison, two ``naive'' procedures, treating the censored data as complete one, are considered including FMV screening filter and QA screening procedure with $\tau=0.3$ (i.e., QAS(0.3)). Another screening utility used is
the Cox-model-based marginal screening procedure (i.e., Cox(SIS)). 

\subsection{Results and conclusions}

Results for the above considered seven experiments are presented in Table 1.
Examination of Table 1 has the following observations.
First, the proposed FMV procedure outperforms all the other screening procedures under the above designed experiments
except for experiment 5 because the MMS values of the FMV procedure are identical to their
true numbers of the active predictors and the FMV procedure has almost the smallest SE value than other screening procedures regardless of the
heavy-tailed responses or outliers. Second, the RCS and FKS procedures behave satisfactory under the considered experiments 1-4
because their corresponding MMS values are identical to
their corresponding true numbers of the active predictors, but they behave unsatisfactory in experiment 5
that involves error heteroscedasticity and the RCS procedure performs poorer than the FKS procedure under error heteroscedasticity assumption.
Third, the considered ten screening methods have the same performance under cases (1a) and (1c)
in experiment 1 in terms of the MMS value. Fourth, the SIRS procedure performs well under cases given in experiments 1 and 4 and (2c) and (2d)
in terms of the MMS value and SE value. Fifth, the NIS, RCS, QAS and SIRS procedures are not directly applicable for experiment 6.
Sixth, the Cox(SIS) procedures do not perform well in experiment 7 because the proportional hazards assumption does not satisfied in experiment 7.

\hfill{$\underline{\overline{\sl Table~1~ about~ here}}$}
\vspace{-3mm}

\section{An example}
\renewcommand{\theequation}{5.\arabic{equation}}
\setcounter{equation}{0}

In this section, the dataset taken from the Boston Housing Study is used to illustrate the proposed FMV screening procedure in R package mlbench.
The dataset is composed of $506$ individuals on $14$ variables.
The key aim of this study is to investigate the effect of clean air on house price.
We take the logarithm of the median value (LMV) of owner occupied homes to be response variable ($Y$),
and other 13 variables to be covariates. These covariates are
per capita crime rate by town (CRIM, $X_1$), proportion of residential land zoned for lots over 25,000 sq.ft (ZN, $X_2$),
proportion of non-retail business acres per town (INDUS, $X_3$), Charles river dummy variable which is $1$
if it is tract bounds river and 0 otherwise (CHAS, $X_4$),
nitric oxides concentration (parts per 10 million, NOX, $X_5$),
average number of rooms per dwelling (RM, $X_6$), proportion of owner-occupied units built prior to 1940 (AGE, $X_7$),
weighted distances to five Boston employment centers (DIS, $X_8$), index of accessibility to radial highways (RAD, $X_9$),
full-value property-tax rate per 10,000 (TAX, $X_{10}$), pupil-teacher ratio by town (OTRATIO, $X_{11}$),
$1000(bk-0.63)^2$ in which bk is the proportion of blacks by town (B, $X_{12}$),
and proportion of population that has a lower status (LSTAT, $X_{13}$). The dataset have even been analyzed by Harrison $\&$ Rubinfeld (1978).

As an illustration of the above proposed FMV screening procedure, we do not make model assumption on $\{Y,X_1,\ldots,X_{13}\}$ but assume that there are 3000 covariates (e.g., $X_1,\ldots,X_{3000}$) in which covariates
$X_1,\ldots,X_{13}$ correspond to the above mentioned 13 covariates, covariates $X_{14},\ldots,X_{91}$ correspond to the interaction effects of any two covariates
(e.g., $X_jX_k$ for $j,k=1,\ldots,13$) among 13 covariates, and covariates $X_{92},\ldots,X_{3000}$ are added to create the noise variables following the Cauchy distribution.
For comparison, we also consider the SIS, SIRS, FKS, NIS and DCS screening procedure.
To evaluate the screening and prediction performance of various methods,
we randomly select $n=350$ individuals out of $506$ individuals for model fitting, and use the rest of the data as the testing set.

First, we examine the performance of various screening procedures by calculating the average number of the selected covariates
in which the above mentioned 91 predictors are included. Here, we consider $K=6$ slice schemes and take $S_k=3,\ldots,\lceil 350^{1/3}\rceil=8$ for $k=1,\ldots,6$, respectively; and slice $Y$ using the $\frac{g}{S_k}$-th sample quantile of $Y$'s for $g=1,\ldots,S_k-1$.
Results for 100 replications are presented in Table 2. It is easily seen from Table 2 that
the FMV, FKS and SIS screening procedures have relatively better performance in selecting the mentioned 91
predictors than other screening procedures.

\hfill{$\underline{\overline{\sl Table~ 2~ about~ here}}$}
\vspace{1mm}

Secondly, we examine how variable screening helps predicting the response variable. Following Ando $\&$ Li (2015), we conduct the model-averaging procedure for high-dimensional regression problems. On the whole, two steps are involved to compute the  mean squared errors (MSE) on the testing sets. The first step is to order the regressors for grouping through utilizing  six screening methods including FMV, SIS, SIRS, FKFS, NIS and DCS.
The second step is to determine the optimal model weights for averaging with a smaller number of regressors.
 For specifying the weights, we construct a class of linear regression candidate models and adopt delete-one cross validation procedure (Ando $\&$ Li, 2015).

We used MSE (averaged squared difference between the observed response $Y$ and the estimated  conditional mean of $Y$) of testing data as the performance measure for each
method. Figure 1 shows the boxplot of MSEs after 100 replicates runs. As
shown in this figure our screening method FMV, combing with model-averaging procedure, yields a nice performance in the sense that it achieves the smallest MSE
median.

\hfill{$\underline{\overline{\sl Figure~ 1~ about~ here}}$}
\vspace{-3mm}

\section{Discussion}

This paper proposes a new model-free screening procedure called as the FMV method for ultrahigh dimensional data analysis.
We further establish its sure screening and rank consistency properties under some wild regularity conditions.
Simulation studies show that the proposed FMV method outperforms the existing screening methods.
An example related to Boston Housing data is used to illustrate the proposed FMV method.

The proposed FMV method has the following merits. First, it is actually robust to model specification (i.e., model free)
and powerful in presence of heavy-tailed distribution assumption on response,
outliers and dependence within covariates.
The introduced slice and fusion steps can be used to deal with many types of responses including discrete, categorical and continuous variables.
Moreover, its sure screening and rank consistency properties are established under some wild regularity conditions, which is conducive
to application with the quantile slice. It is interesting to address the optimal slice scheme.
Second, we present two steps to explore the accuracy of predictions for various screening methods in analyzing Boston Housing data.
The first step is to order predictors using model-free screening procedure,
and the second step is to fit candidate model. However, it is interesting and preferred to propose
a model-free and robust model average procedure for studying the accuracy of predictions after a model-free variable screening method, which is out of scope of this work.
Third, some works on boosting the performance of model-based screening methods have been done by adopting an iterative screening and model-fitting procedure.
However, we have not derived a similar iterative procedure for model-free screening method. This is an open problem left for further research.

\section*{Appendix}
\renewcommand{\theequation}{A.\arabic{equation}}
\setcounter{equation}{0}
\noindent  \emph{\bf Proof of Lemma 1}
We start to prove the first conclusion. If $X_j$ is independent of $Y$, then $X_j$ will be independent of any $G$, which is a function of $Y$. Therefore, $MV_j^{\mathbb{S}}=0$ for all $\mathbb{S}$. Now suppose $MV_j^{\mathbb{S}}=0$ for all choices of $\mathbb{S}$. For any $y$, consider $G=1$ if $Y\leq y$ and $G=2$ otherwise. Because $MV_j^{\mathbb{S}}=0$, $X_j$ is independent of $G$. Consequently, $\mathbb{P}(Y\leq y|X_j)=\mathbb{P}(Y\leq y)$ for all $y$, and $Y$ is independent of $X_j$.

Secondly, suppose there exists $\mathbb{S}$ such that $MV_j^{\mathbb{S}}=0$. Then $X_j\perp G$ for the corresponding $G$. Therefore, $\mathbb{P}(a_1\leq Y< a_2|X_j)=\mathbb{P}(Y< a_2|X_j)-\mathbb{P}(Y< a_1|X_j)=\mathbb{P}(G=1|X_j)=\mathbb{P}(G=1)$ is a constant, which contradicts our assumptions. Thus we must have $MV_j^{\mathbb{S}}\neq 0$.

Now, we turn to the third conclusion. Because $X_j$ is not independent of $Y$, $MV_j^{\mathbb{S}}>0$. Hence it suffices to show that $MV_j^{\mathbb{S}}\rightarrow MV_j$ as $S\rightarrow\infty.$

The certification of Lemma 1(iii) is equivalent to prove
$$\sum_{g=1}^Sp_g^\mathbb{S}\int [F_j^\mathbb{S}(x|G=g)-F_j(x)]^2dF_j(x)\rightarrow\iint[F_j(x|Y=y)-F_j(x)]^2dF_j(x)dF_Y(y).$$
We just prove that
$$\sum_{g=1}^Sp_g^\mathbb{S}[F_j^\mathbb{S}(x|G=g)-F_j(x)]^2\rightarrow\int[F_j(x|Y=y)-F_j(x)]^2dF_Y(y).$$

Because $F_j(x|y)$ is continuous in $y$, for any $\epsilon$, there exists $\delta>0$, such that $\sup\limits_{x\in\mathbb{R}_{X_j}}|F_j(x|y)-F_j(x|y^*)|<\epsilon$ for any fixed $x$ and $|y-y^*|<\delta$. Take $\phi=\mathbb{P}(\{y:|y-y^*|<\delta\})$, because $\max\limits_{g=1,\ldots,S}\mathbb{P}(G=g)\rightarrow 0$, there exists $S^*$ such that $\mathbb{P}(G=g)<\frac{\phi}{2}$ when $S>S^*$. In such cases, there exists $g$, $[a_g,a_{g+1})\subset(y^*-\delta,y^*+\delta),$ meaning $|y^*-a_{g}|<\delta$ and $|F_j^\mathbb{S}(x|G=g)-F_j(x|y^*)|<\epsilon$.
 So when $S>S^*$,
\begin{eqnarray*}
&&\sum_{g=1}^Sp_g^\mathbb{S}[F_j^\mathbb{S}(x|G=g)-F_j(x)]^2=\sum_{g=1}^Sp_g^\mathbb{S}[F_j^\mathbb{S}(x|G=g)-F_j(x|y_g^*)+F_j(x|y_g^*)-F_j(x)]^2\\
&=&\sum_{g=1}^Sp_g^\mathbb{S}[F_j^\mathbb{S}(x|G=g)-F_j(x|y_g^*)]^2-2\sum_{g=1}^Sp_g^\mathbb{S}[F_j^\mathbb{S}(x|G=g)-F_j(x|y_g^*)][F_j(x|y_g^*)-F_j(x)]\\
&&+\sum_{g=1}^Sp_g^\mathbb{S}[F_j(x|y_g^*)-F_j(x)]^2\\
&=&H_{j1}+H_{j2}+H_{j3}.
\end{eqnarray*}

Note $H_{j1}<\epsilon^2$, $H_{j2}<2\epsilon$. For $H_{j3}$, we use the Condition $\lim\limits_{S\rightarrow\infty} p_g^\mathbb{S}S=1$, meaning there exists $S^*$, $|p_g^\mathbb{S}-1/S|<\epsilon/S$ for $S>S^*$.  Set
\[
\begin{array}{llll}
|\Delta_{H_{j3}}|&=&|\sum_{g=1}^Sp_g^\mathbb{S}[F_j(x|y_g^*)-F(x)]^2-\sum_{g=1}^S\frac{1}{S}[F_j(x|a_g)-F_j(x)]^2|\\
&\leq&|\sum_{g=1}^Sp_g^\mathbb{S}[F_j(x|y_g^*)-F(x)]^2-\sum_{g=1}^S\frac{1}{S}[F_j(x|y_g^*)-F_j(x)]^2|\\
&&+|\sum_{g=1}^S\frac{1}{S}[F_j(x|y_g^*)-F(x)]^2-\sum_{g=1}^S\frac{1}{S}[F_j(x|a_g)-F_j(x)]^2|\\
&\leq&\sum_{g=1}^S|p_g^\mathbb{S}-\frac{1}{S}|[F_j(x|y_g^*)-F_j(x)]^2+2\sum_{g=1}^S\frac{1}{S}\sup\limits_{x\in\mathbb{R}_{X_j}}|[F_j(x|y_g^*)-F_j(x|a_g)]|\\
&\leq&\epsilon+2\epsilon=3\epsilon.
\end{array}
\]
Therefore,
\[
\begin{array}{llll}
\mathbb{P}(|\sum_{g=1}^Sp_g^\mathbb{S}[F_j(x|y_g^*)-F(x)]^2-\int[F_j(x|y)-F_j(x)]^2dF_Y(y)|\ge 6\epsilon)\\
\leq \mathbb{P}(|\Delta_{H_{j3}}|\ge 3\epsilon)+\mathbb{P}(|\sum_{g=1}^S\frac{1}{S}[F_j(x|a_g)-F_j(x)]^2-E_{a_g}[F_j(x|a_g)-F(x)]^2|\ge 3\epsilon)\\
\leq \sup\limits_{x\in\mathbb{R}_{X_j}}\int[F_j(x|y)-F_j(x)]^4dF_Y(y)/(9S\epsilon^2)
\end{array}
\]
So we finish the certification of Lemma 1.

Define
\[
\begin{array}{llll}
p_g^\mathbb{S}=P\{G=g\};\\
h_j^\mathbb{S}=\sum\limits_{g=1}^{S}I\{G=g\}\int [F_j^\mathbb{S}(x|G=g)-F_j(x)]^2dF_j(x);\\
h_j^\mathbb{S}(g,x)=[F_j^\mathbb{S}(x|G=g)-F_j(x)]^2;\\
\Delta^\mathbb{S}(g)=I\{G=g\};\\
\Delta_{j}(x)=I\{X_j\leq x\};\\
\Delta_j^\mathbb{S}(g,x)=I\{X_j\leq x,G=g\},\\
h_{ij}^\mathbb{S}=\sum\limits_{g=1}^{S}I\{a_g\leq Y_i<a_{g+1}\}\int [F_j^\mathbb{S}(x|G=g)-F_j(x)]^2dF_j(x);\\ \Delta_i^\mathbb{S}(g)=I\{a_g\leq Y_i<a_{g+1}\};\\ \Delta_{ij}(x)=I\{X_{ij}\leq x\};\\
\Delta_{ij}^\mathbb{S}(g,x)=I\{X_{ij}\leq x,a_g\leq Y_i<a_{g+1}\}.
\end{array}
\]
where $\{G=g\}=\{a_g\leq Y<a_{g+1}\}$.

\begin{lemma}
For  slice scheme $\mathbb{S}$, $\epsilon\in(0,1)$ and $1\leq g\leq S$, the following inequalities are valid for univariate $X_j$.
\begin{eqnarray}\label{lemma21}
&&\mathbb{P}\Big{\{}\Big|1/n\sum_{i=1}^nh_j^\mathbb{S}(g,X_{ij})-Eh_j^\mathbb{S}(g,X_j)\Big|\geq \epsilon\Big{\}}\leq 2\exp\{-2n\epsilon^2\};\\\label{lemma22}
&&\mathbb{P}\Big{\{}\Big|1/n\sum_{i=1}^nh_{ij}^\mathbb{S}-Eh_j^\mathbb{S}\Big|\geq \epsilon\Big{\}}\leq 2\exp\{-2n\epsilon^2\};\\\label{lemma23}
&&\mathbb{P}\Big{\{}\Big|1/n\sum_{i=1}^n\Delta_i^\mathbb{S}(g)-E\Delta^\mathbb{S}(g)\Big|\geq \epsilon\Big{\}}\leq 2\exp\Big{\{}-\frac{n\epsilon^2}{2(p_g^\mathbb{S}+\epsilon/3)}\Big{\}};\\\label{lemma24}
&&\mathbb{P}\Big{\{}\sup_{x\in{\mathbb{R}_{X_j}}}\Big|1/n\sum_{i=1}^n\Delta_{ij}(x)-E\Delta_j(x)\Big|\geq \epsilon\Big{\}}\leq 2(n+1)\exp\{-2n\epsilon^2\};\\\nonumber
&&\mathbb{P}\Big{\{}\sup_{x\in{\mathbb{R}_{X_j}}}\Big|1/n\sum_{i=1}^n\Delta_{ij}^\mathbb{S}(g,x)-E\Delta_j^\mathbb{S}(g,x)\Big|\geq \epsilon\Big{\}}\leq 2(n+1)\exp\Big{\{}-\frac{n\epsilon^2}{2(p_g^\mathbb{S}+\epsilon/3)}\Big{\}}.\\\label{lemma25}
&&
\end{eqnarray}
\end{lemma}

\noindent  \emph{\bf Proof of Lemma 2}
Since $|h_j^\mathbb{S}(g,X_{ij})|=[F_j^\mathbb{S}(X_{ij}|G=g)-F_j(X_{ij})]^2\leq 1$ and $|h_{ij}^\mathbb{S}|\leq 1$, here we use the~Hoeffding's~ inequality (Cui et al., 2015) to obtain the inequalities (\ref{lemma21}) and (\ref{lemma22}). For (\ref{lemma23}), $\Delta_i^\mathbb{S}(g)\sim Bernoulli(p_g^\mathbb{S})$ with $E\Delta_i^\mathbb{S}(g)=p_g^\mathbb{S}$ and $\Delta_1^\mathbb{S}(g)+\ldots+\Delta_n^\mathbb{S}(g)\sim Binomial(n,p_g^\mathbb{S})$, which implies $Var(\Delta_1^\mathbb{S}(g)+\ldots+\Delta_n^\mathbb{S}(g))=np_g^\mathbb{S}(1-p_g^\mathbb{S})\leq np_g^\mathbb{S}$ and $|\Delta_i^\mathbb{S}(g)-p_g^\mathbb{S}|\leq 1$. Thus by~Bernstein's~inequality  (Cui et al., 2015), we have
\begin{eqnarray*}
\mathbb{P}\Big{\{}\Big|1/n\sum_{i=1}^n\Delta_i^\mathbb{S}(g)-E\Delta^\mathbb{S}(g)\Big|\geq\epsilon\Big{\}}&=&\mathbb{P}\Big{\{}\Big|\sum_{i=1}^n(\Delta_i^\mathbb{S}(g)-p_g^\mathbb{S})\Big|\geq n\epsilon\Big{\}}\leq 2\exp\Big{\{}-\frac{n^2\epsilon^2}{2(np_g^\mathbb{S}+{n\epsilon}/3)}\Big{\}}\\
&\leq&2\exp\Big{\{}-\frac{n\epsilon^2}{2(p_g^\mathbb{S}+\epsilon/3)}\Big{\}}.
\end{eqnarray*}
Note that $|\Delta_{ij}(x)-E\Delta_j(x)|=|I\{X_{ij}\leq x\}-F_j(x)|\leq 1$, then we apply~Hoeffding's~inequality and empirical process theory~(Pollard, 1984)~to obtain (\ref{lemma24}). At last, $|\Delta_{ij}^\mathbb{S}(g,x)-E\Delta_j^\mathbb{S}(g,x)|=|I\{X_{ij}\leq x, a_g\leq Y_i<a_{g+1}\}-F_j^\mathbb{S}(x|G=g)p_g^\mathbb{S}|\leq 1|$, then we apply~Bernstein's~inequality and empirical process theory~to obtain (\ref{lemma25}).
\begin{lemma}
$\widehat{a}_g=\widehat{F}_Y^{-1}(\frac{g-1}{S})$ is the sample quantile for $Y$, where $\widehat{F}_Y(y)$ is the empirical distribution of $F_Y(y)$. Then with a probability greater than $1-C\exp(-C\frac{n}{S^2})$ and $1- C\exp(-C\frac{n\Delta^2_\mathcal{E}}{S^2})$, respectively, we have
\begin{eqnarray}\label{lemma31}
3/(4S)\leq\mathbb{P}(\widehat{a}_g\leq Y<\widehat{a}_{g+1})\leq 5/(4S),\\\label{lemma32}
\mathbb{P}(\widehat{a}_g\leq Y<\widehat{a}_{g+1})\leq(2+\Delta_\mathcal{E})/2S.
\end{eqnarray}
\end{lemma}\noindent  \emph{\bf Proof of Lemma 3}
Firstly, under event $A=\sup_y|\widehat{F}_Y(y)-F_Y(y)|\leq\frac{1}{8S}$, then  we conclude (\ref{lemma31}). Because, under event A,
\[
\begin{array}{llll}
\mathbb{P}(\widehat{a}_g\leq Y<\widehat{a}_{g+1})&=&\mathbb{P}(\frac{g-1}{S}\leq\widehat{F}_Y(Y)<\frac{g}{S})\\
&\ge& \mathbb{P}(\frac{g-1}{S}+\frac{1}{8S}\leq F_Y(Y)<\frac{g}{S}-\frac{1}{8S})=\frac{3}{4S}\\
{\rm On\;the\;other\;hand,}&\leq& \mathbb{P}(\frac{g-1}{S}-\frac{1}{8S}\leq F_Y(Y)<\frac{g}{S}+\frac{1}{8S})=\frac{5}{4S}.
\end{array}
\]
(\ref{lemma32}) is concluded from  event $B=\sup_y|\widehat{F}_Y(y)-F_Y(y)|\leq\frac{\Delta_\mathcal{E}}{4S}$, because
 \[
\begin{array}{llll}
&&\mathbb{P}(\widehat{a}_g\leq Y<\widehat{a}_{g+1})\\
&=&\mathbb{P}(\frac{g-1}{S}\leq\widehat{F}_Y(Y)<\frac{g}{S})\\
&\leq& \mathbb{P}(\frac{g-1}{S}-\frac{\Delta_\mathcal{E}}{4S}\leq F_Y(Y)<\frac{g}{S}+\frac{\Delta_\mathcal{E}}{4S})=\frac{2+\Delta_\mathcal{E}}{2S}\leq \frac{1+\Delta_\mathcal{E}}{S}.
 \end{array}
\]
Then by the Dvoretzky-Kiefer-Wolfowitz
inequality,  we have $\mathbb{P}(A)\ge 1-C\exp(-C
\frac{n}{S^2} )$ and $\mathbb{P}(B)\ge 1-C\exp(-C
\frac{n\Delta_\mathcal{E}^2}{S^2} )$.
Therefore, we have Lemma 3.

\begin{lemma}
{\rm (i)}
$$\mathbb{P}(|\widehat{\omega}^\circ_j-\omega^\circ_j|\ge K\frac{\Delta_{\mathcal{E}}}{2})
\le CK\exp\{-Cn^{1-\kappa}\Delta_{\mathcal{E}}^2+(1+\kappa)\log(n)\}.$$
{\rm (ii)}
$$\mathbb{P}(|\widehat{\omega}_j-\omega_j|\ge K\frac{\Delta_{\mathcal{E}}}{2})
\le CK\exp\{-Cn^{1-\kappa}\Delta_{\mathcal{E}}^2+(1+\kappa)\log(n)\}.$$
{\rm (iii)} Under Condition (C2), we have

$$ \mathbb{P}\{|\omega_j-\omega_j^\circ|\geq K\frac{\Delta_{\mathcal{E}}}{2}\}
  \leq CK\big(\exp\{-Cn^{1-2\kappa}\Delta_{\mathcal{E}}^2\}\big).$$

\end{lemma}
\noindent  \emph{\bf Proof of Lemma 4}
Define
\begin{eqnarray*}
  \widehat{\omega}^\circ_j(S)&=&1/n\sum_{i=1}^n\sum_{g=1}^{S}\widehat{p}_g^\mathbb{S}[\widehat{F}_j^\mathbb{S}(x_i|G=g)-\widehat{F}_j(x_i)]^2
  =\sum_{g=1}^{S}\widehat{p}_g^\mathbb{S}\int[\widehat{F}_j^\mathbb{S}(x|G=g)-\widehat{F}_j(x)]^2 d\widehat{F}_j(x),\\
  \widetilde{\omega}^\circ_j(S)&=&\sum_{g=1}^{S}\widehat{p}_g^\mathbb{S}\int[F_j^\mathbb{S}(x|G=g)-{F}_j(x)]^2 d{F}_j(x),\\
  {\omega}^\circ_j(S)&=&\sum_{g=1}^{S}p_g^\mathbb{S}\int[F_j^\mathbb{S}(x|G=g)-{F}_j(x)]^2 d{F}_j(x).
\end{eqnarray*}
Then, we have
\begin{eqnarray*}
  &&\widehat{\omega}^\circ_j(S)- {\omega}^\circ_j(S)\\
  &=&\widehat{\omega}^\circ_j(S)-\widetilde{\omega}^\circ_j(S)+\widetilde{\omega}^\circ_j(S)+ {\omega}^\circ_j(S)\\
   &=&\sum_{g=1}^{S}\widehat{p}_g^\mathbb{S}\left\{\int[\widehat{F}_j^\mathbb{S}(x|G=g)-\widehat{F}_j(x)]^2 d\widehat{F}_j(x)-\int[F_j^\mathbb{S}(x|G=g)-{F}_j(x)]^2 d{F}_j(x)\right\}\\
   &&+\sum_{g=1}^{S}(\widehat{p}_g^\mathbb{S}-p_g^\mathbb{S})\int[F_j^\mathbb{S}(x|G=g)-{F}_j(x)]^2 d{F}_j(x)\\
   &=&\sum_{g=1}^{S}\widehat{p}_g^\mathbb{S}\int\left\{[\widehat{F}_j^\mathbb{S}(x|G=g)-\widehat{F}_j(x)]^2 -[F_j^\mathbb{S}(x|G=g)-{F}_j(x)]^2 \right\}d\widehat{F}_j(x)\\
   &&+\sum_{g=1}^{S}\widehat{p}_g^\mathbb{S}\int[F_j^\mathbb{S}(x|G=g)-{F}_j(x)]^2 d[\widehat{F}_j(x)-{F}_j(x)]\\
   &&+\sum_{g=1}^{S}(\widehat{p}_g^\mathbb{S}-p_g^\mathbb{S})\int[F_j^\mathbb{S}(x|G=g)-{F}_j(x)]^2 d{F}_j(x)\\
   &&:=I_{j1}+I_{j2}+I_{j3}
\end{eqnarray*}
For $I_{j1}$.
\begin{eqnarray*}
  |I_{j1}|&&\leq 2\underset{g}{\max}\int |[\widehat{F}_j^\mathbb{S}(x|G=g)-F_j^\mathbb{S}(x|G=g)]-[\widehat{F}_j(x)-{F}_j(x)]|d\widehat{F}_j(x)\\
  &&\leq 2\underset{g}{\max}\underset{x\in \mathbb{R}_{X_j}}{\sup}(|[\widehat{F}_j^\mathbb{S}(x|G=g)-F_j^\mathbb{S}(x|G=g)]|+|[\widehat{F}_j(x)-{F}_j(x)]|)\\
  &&:=J_{j1}+J_{j2}
\end{eqnarray*}
where the first inequality holds by
$\sum_{g=1}^{S}\widehat{p}_g^\mathbb{S}=1 $ and
  $|[\widehat{F}_j^\mathbb{S}(x|G=g)-F_j^\mathbb{S}(x|G=g)]+[\widehat{F}_j(x)-{F}_j(x)]|
  \leq |[\widehat{F}_j^\mathbb{S}(x|G=g)-F_j^\mathbb{S}(x|G=g)]|+|[\widehat{F}_j(x)-{F}_j(x)]|\leq2$
and the second inequality is derived by $\int d\widehat{F}_j(x)=1$.

Then we control the term $J_{j1}$.
\begin{eqnarray*}
  J_{j1}&=&\underset{g}{\max}\underset{x\in \mathbb{R}_{X_j}}{\sup}|[\widehat{F}_j^\mathbb{S}(x|G=g)-F_j^\mathbb{S}(x|G=g)]|\\
  &=&\underset{g}{\max}\underset{x\in \mathbb{R}_{X_j}}{\sup}|1/n\sum_{i=1}^{n}\frac{\Delta_{ij}^\mathbb{S}(g,x)}{\widehat{p}_g^\mathbb{S}}- \frac{E\Delta_j^\mathbb{S}(g,x)}{p_g^\mathbb{S}}|\\
  &\leq& \underset{g}{\max}\underset{x\in \mathbb{R}_{X_j}}{\sup}\left(\frac{|1/n\sum_{i=1}^{n}\Delta_{ij}^\mathbb{S}(g,x)-E\Delta_j^\mathbb{S}(g,x)|}{\widehat{p}_g^\mathbb{S}}
  +\frac{E\Delta_j^\mathbb{S}(g,x)|\widehat{p}_g^\mathbb{S}-p_g^\mathbb{S}|}{\widehat{p}_g^\mathbb{S}p_g^\mathbb{S}}\right)\\
  &=&\underset{g}{\max}\underset{x\in \mathbb{R}_{X_j}}{\sup}\frac{|1/n\sum_{i=1}^{n}\Delta_{ij}^\mathbb{S}(g,x)-E\Delta_j^\mathbb{S}(g,x)|}{\widehat{p}_g^\mathbb{S}}
  +\underset{g}{\max}\frac{E\Delta_j^\mathbb{S}(g,x)|\widehat{p}_g^\mathbb{S}-p_g^\mathbb{S}|}{\widehat{p}_g^\mathbb{S}p_g^\mathbb{S}}
\end{eqnarray*}
where the equality holds due to $\underset{x\in \mathbb{R}_{X_j}}{\sup}E \Delta_j^\mathbb{S}(g,x)=\underset{x\in \mathbb{R}_{X_j}}{\sup}\mathbb{P}(X_j\leq x,G=g)=p_g^\mathbb{S}$.
Thus,  for any $0\leq \epsilon < \frac{1}{2}$,
\begin{eqnarray}
  &&\mathbb{P}(J_{j1}\geq \epsilon)\nonumber \\\label{probab1}
 &\leq& \mathbb{P}\left\{\left(\underset{g}{\max}\underset{x\in \mathbb{R}_{X_j}}{\sup}\frac{|1/n\sum_{i=1}^{n}\Delta_{ij}^\mathbb{S}(g,x)-E\Delta_j^\mathbb{S}(g,x)|}{\widehat{p}_g^\mathbb{S}}
  +\underset{g}{\max}\frac{E\Delta_j^\mathbb{S}(g,x)|\widehat{p}_g^\mathbb{S}-p_g^\mathbb{S}|}{\widehat{p}_g^\mathbb{S} p_g^\mathbb{S}}\right)\geq\epsilon\right\}\nonumber  \\
&\leq& \mathbb{P}(\underset{g}{\max}\underset{x\in \mathbb{R}_{X_j}}{\sup}{|1/n\sum_{i=1}^{n}\Delta_{ij}^\mathbb{S}(g,x)-E\Delta_j^\mathbb{S}(g,x)|\geq
 \frac{\epsilon}{2S}})\nonumber \\\label{probab2}
 &&+\mathbb{P}(\underset{g}{\max}|1/n\sum_{i=1}^n{\Delta_i^\mathbb{S}(g)-E \Delta_{g}^\mathbb{S}}|\geq\frac{\epsilon}{2S}) \\
 & \leq& 2(n+1)S \exp\left\{-\frac{n(\epsilon/2S)^2}{2(p_g^\mathbb{S}+\frac{\epsilon}{6S})}\right\}
 +S\exp\left\{-\frac{n(\epsilon/2S)^2}{2(p_g^\mathbb{S}+\frac{\epsilon}{6S})}\right\}\nonumber \\
 &\leq &(2n+3)S \exp\{-C\frac{n\epsilon^2}{S}\nonumber \}
\end{eqnarray}
for some constant $C>0$, where the second inequality holds
because  $\underset{g}{\max}|\widehat{p}_g^\mathbb{S}-p_g^\mathbb{S}|=\underset{g}{\max}|1/n\sum_{i=1}^n{\Delta_i^\mathbb{S}(g)-E \Delta^\mathbb{S}(g)}|.$ 
Then we apply inequalities (\ref{lemma24}), (\ref{lemma21}), (\ref{lemma22}) in Lemma 2 and obtain that
$$ \mathbb{P}(J_{j2}\geq\epsilon)=\mathbb{P}\{\underset{x\in \mathbb{R}_{X_j}}{\sup}|\widehat{F}_j(x)-{F}_j(x)|\geq\epsilon\}
  \leq 2(n+1)\exp(-2n\epsilon^2).$$
So, for some constant $C$,
 \begin{equation}{\label{Ij1}}
 \mathbb{P}(I_{j1}\geq\epsilon)\leq CnS\exp(-C\frac{n\epsilon^2}{S}).
\end{equation}
\begin{equation}{\label{Ij2}}
\mathbb{P}(I_{j2}\geq\epsilon)=\mathbb{P}\{|\sum_{g=1}^{S}\widehat{p}_g^\mathbb{S}(1/n\sum_{i=1}^{n}h_j^\mathbb{S}(g,X_{ij})-E h_j^\mathbb{S}(g,X_j))|\geq\epsilon\}
  \leq 2S\exp(-2n\epsilon^2).
\end{equation}
\begin{equation}{\label{Ij3}}
  \mathbb{P}\{|I_{j3}|\geq \epsilon\} =\mathbb{P}\{|1/n\sum_{i=1}^{n}h_{ij}^\mathbb{S}-Eh_j^\mathbb{S}|\geq\epsilon\}\leq 2\exp(-2n\epsilon^2).
\end{equation}

Under the equations (\ref{Ij1})-(\ref{Ij3}), we have 
\begin{equation*}
  \mathbb{P}\{|\widehat{\omega}_j^\circ(S)-{\omega}_j^\circ(S)|\geq \epsilon\}\leq CnS\exp\{-C\frac{n\epsilon^2}{S}\}
\end{equation*}
leading to

\begin{eqnarray*}
  \mathbb{P}\{|\widehat{\omega}_j^\circ-\omega_j^\circ|\geq K\frac{\Delta_{\mathcal{E}}}{2}\}
  &\leq&\mathbb{P}\{\sum\limits_{k=1}^K|\widehat{\omega}_j^\circ(S_k)-{\omega}_j^\circ(S_k)|\geq K\frac{\Delta_{\mathcal{E}}}{2}\}\\
  &\leq& \sum\limits_{k=1}^{K}\mathbb{P}\{|\widehat{\omega}_j^\circ(S_k)-{\omega}_j^\circ(S_k)|\geq \frac{\Delta_{\mathcal{E}}}{2}\}\\
  &=& CK\exp\{-Cn^{1-\kappa}\Delta_{\mathcal{E}}^2+(1+\kappa)\log(n)\},
\end{eqnarray*}
for $S_k=O(n^\kappa)$. 
Similarly, we can show Lemma 4 (ii) under (\ref{lemma31}).
For proving Lemma 4 (iii), we define
\[
\begin{array}{llll}
&\omega_j(S)=\sum_{g=1}^{S}p_g^{\widehat{\mathbb{S}}}\int[F_j^{\widehat{\mathbb{S}}}(x|\widehat{G}=g)-F_j(x)]^2dF_j(x),\\
&\omega_j^{\circ}(S)=\sum_{g=1}^{S}p_g^\mathbb{S}\int[F_j^\mathbb{S}(x|G=g)-F_j(x)]^2dF_j(x),\\
&\omega_j^*(S)=\sum_{g=1}^{S}p_g^{\widehat{\mathbb{S}}}\int[F_j^\mathbb{S}(x|G=g)-F_j(x)]^2dF_j(x),
 \end{array}
\]
where $\{\widehat{G}=g\}=\{\widehat{a}_g\leq Y<\widehat{a}_{g+1}\}$, $p_g^{\widehat{\mathbb{S}}}=\mathbb{P}\{\widehat{G}=g\},$ then we have
\[
\begin{array}{llll}
\omega_j(S)-\omega_j^{\circ}(S)  &=&  \omega_j(S)-\omega_j^*(S)+\omega_j^*(S)-\omega_j^{\circ}(S)\\
&=& \sum_{g=1}^{S}p_g^{\widehat{\mathbb{S}}}\{\int[F_j^{\widehat{\mathbb{S}}}(x|\widehat{G}=g)-F_j(x)]^2dF_j(x)-\int[F_j^\mathbb{S}(x|G=g)-F_j(x)]^2dF_j(x)\}\\
&+& \sum_{g=1}^{S}(p_g^{\widehat{\mathbb{S}}}-p_g^\mathbb{S})\int[F_j^\mathbb{S}(x|G=g)-F_j(x)]^2dF_j(x)\\
&=& L_{j1}+L_{j2}.
\end{array}
\]
among which,
\[
\begin{array}{llll}
L_{j1}&\le& 2\max_g\sup_{x\in \mathbb{R}_{X_j}}|F_j^{\widehat{\mathbb{S}}}(x|\widehat{G}=g)-F_j^\mathbb{S}(x|G=g)|\\
&\le& 2\max_g\sup_{x\in \mathbb{R}_{X_j}}|F_j^{\widehat{\mathbb{S}}}(x|\widehat{G}=g)-F_j(x|y^*)|+2\max_g\sup_{x\in \mathbb{R}_{X_j}}|F_j(x|y^*)-F_j^\mathbb{S}(x|G=g)|\\
&\leq&2\times \frac{\Delta_{\mathcal{E}}}{8}+2\times \frac{\Delta_{\mathcal{E}}}{8}= \frac{\Delta_{\mathcal{E}}}{2}.
\end{array}
\]
where $y^*\in \{\widehat{G}=g\}\bigcap\{G=g\}$. Under Condition (C2) and (\ref{lemma32}), $\max_g |p_g^{\widehat{\mathbb{S}}}-p_g^\mathbb{S}|\leq\frac{\Delta_\mathcal{E}}{2S},$
we have
\begin{equation}
L_{j2}\le S\max_g |p_g^{\widehat{\mathbb{S}}}-p_g^\mathbb{S}|\leq\frac{\Delta_\mathcal{E}}{2}.
\end{equation}
Therefore, we have

\begin{eqnarray*}
  \mathbb{P}\{|\omega_j-\omega_j^\circ|\geq K\frac{\Delta_{\mathcal{E}}}{2}\}
  &\leq&\mathbb{P}\{\sum\limits_{k=1}^K|\omega_j(S_k)-{\omega}_j^\circ(S_k)|\geq K\frac{\Delta_{\mathcal{E}}}{2}\}\\
  &\leq& \sum\limits_{k=1}^{K}\mathbb{P}\{|\omega_j(S_k)-{\omega}_j^\circ(S_k)|\geq \frac{\Delta_{\mathcal{E}}}{2}\}\\
  &=& CK\big(\exp\{-Cn^{1-2\kappa}\Delta_{\mathcal{E}}^2\}\big). \end{eqnarray*}

\noindent  \emph{\bf Proof of Theorem 1}
Under Lemma 4 (ii) and (iii), we can conclude Theorem 1 (i) from $|\widehat{\omega}_j-\omega_j^\circ|\leq |\widehat{\omega}_j-\omega_j|+|\omega_j-\omega_j^\circ|$.

We turn to  Theorem 1 (ii). Under Lemma 4 (ii), if $\max\limits_{j}|\widehat{\omega}_j^{\circ}-\omega_j^{\circ}|< K\frac{\Delta_\mathcal{E}}{2}$, we must have $\mathcal{D}\subset\widehat{\mathcal{D}}_{(oracle)}$. This is indeed true, because combining it with Condition (C1), we have
\[
\begin{array}{llll}
&\widehat{\omega}_j^{\circ}> \omega_j^{\circ}-K\frac{\Delta_\mathcal{E}}{2} \geq \min_{j\in \mathcal{E}}\omega_j^{\circ}-K\frac{\Delta_\mathcal{E}}{2},\;for\;j\in \mathcal{E},\\
&\widehat{\omega}_j^{\circ}<\omega_j^{\circ}+K\frac{\Delta_\mathcal{E}}{2} \leq \max_{j\notin \mathcal{E}}\omega_j^{\circ}+K\frac{\Delta_\mathcal{E}}{2},\;for \;j\notin\mathcal{E}.
\end{array}
\]
Hence,
\[
\begin{array}{llll}
&\mathbb{P}(\mathcal{D}\subset \widehat{\mathcal{D}}_{(oracle)})\ge \mathbb{P}(\mathcal{E}\subset \widehat{\mathcal{D}}_{(oracle)})\ge \mathbb{P}(\max\limits_{j}|\widehat{\omega}_j^{\circ}-\omega_j^{\circ}|< K\frac{\Delta_\mathcal{E}}{2})\\
&=  1-\mathbb{P}(\max\limits_{j}|\widehat{\omega}_j^{\circ}-\omega_j^{\circ}| \ge K\frac{\Delta_\mathcal{E}}{2})\\
&\ge 1-CKp\exp\{-Cn^{1-\kappa}\Delta_{\mathcal{E}}^2+(1+\kappa)\log(n)\}.
\end{array}
\]
For Theorem 1 (iii), we again have that, if $\max\limits_j|\widehat{\omega}_j-\omega_j^{\circ}| < K\Delta_\mathcal{E}/2$, we must have $\mathcal{D}\subset\widehat{\mathcal{D}}$.
Under Theorem 1 (i), we have
\[
\begin{array}{llll}
&\mathbb{P}(\mathcal{D}\subset \widehat{\mathcal{D}})\ge \mathbb{P}(\mathcal{E}\subset \widehat{\mathcal{D}})\ge \mathbb{P}(\max\limits_{j}|\widehat{\omega}_j-\omega_j^{\circ}|< K\Delta_\mathcal{E}/2)\\
&= 1-\mathbb{P}(\max\limits_{j}|\widehat{\omega}_j-\omega_j^\circ|\ge K\frac{\Delta_\mathcal{E}}{2})\\
&\ge 1-CKp\exp\{-Cn^{1-\kappa}\Delta_{\mathcal{E}}^2+(1+\kappa)\log(n)\}.
\end{array}
\]

\noindent  \emph{\bf Proof of Theorem 2}
\begin{eqnarray*}
   \mathbb{P}\{(\underset{j\in \mathcal{D}}{\min}{\widehat{\omega}_j}-\underset{j\notin \mathcal{D}}{\max}{\widehat{\omega}_j})<K\frac{\Delta_\mathcal{E}}{2}\}
  &\leq& \mathbb{P}\{(\underset{j\in \mathcal{D}}{\min}\widehat{\omega}_j-\underset{j\notin \mathcal{D}}{\max}{\widehat{\omega}_j})-(\underset{j\in \mathcal{D}}{\min}\omega_j^\circ-\underset{j\notin \mathcal{D}}{\max}\omega_j^\circ)\leq -K\frac{\Delta_\mathcal{E}}{2}\}\\
  &\leq& \mathbb{P}\{|(\underset{j\in \mathcal{D}}{\min}\widehat{\omega}_j-\underset{j\notin \mathcal{D}}{\max}{\widehat{\omega}}_j)-(\underset{j\in \mathcal{D}}{\min}\omega_j^\circ-\underset{j\notin \mathcal{D}}{\max}\omega_j^\circ)|\geq K\frac{\Delta_\mathcal{E}}{2}\}\\
  &\leq& \mathbb{P}\{2\underset{1\leq j \leq p}{\max}|\widehat{\omega}_j-\omega_j^\circ|\geq K\frac{\Delta_\mathcal{E}}{2}\}\\
  &\leq& C\sum_{k=1}^{K}pnS_k\exp\{-C\frac{n\Delta^2_\mathcal{E}}{S_k}\}\\
  &\leq& CKp\exp\{-Cn^{1-\kappa}\Delta_{\mathcal{E}}^2+(1+\kappa)\log(n)\},
\end{eqnarray*}
for some constant $C>0$, where the first inequality follows Condition (C1), and the last inequality is implied by Theorem 1 (i).

Because $\frac{\log pK}{n^{1-\kappa}}=o(\Delta^2_\mathcal{E})$ and $n^{2\kappa-1}=o(\Delta^2_\mathcal{E})$ imply that $p\leq \frac{1}{K}\exp\{\frac{Cn^{1-\kappa}\Delta^2_\mathcal{E}}{2}\}$, and $\frac{Cn^{1-\kappa}\Delta^2_\mathcal{E}}{2}\geq 4\log n$, $(1+\kappa)\log(n)\leq 2\log n$ for large $n$. Then we have for some $n_0$,
\begin{eqnarray*}
\sum_{n=n_0}^{\infty}n^{1+\kappa}pK\exp\{-Cn^{1-\kappa}\Delta^2_\mathcal{E}\}&\leq& \sum_{n=n_0}^{\infty}\exp\{(1+\kappa)\log n+\frac{Cn^{1-\kappa}\Delta^2_\mathcal{E}}{2}-Cn^{1-\kappa}\Delta^2_\mathcal{E} \}\\
&\leq&\sum_{n=n_0}^{\infty}\exp\{(1+\kappa)\log n-4\log n\}
\leq \sum_{n=n_0}^{\infty}n^{-2}<\infty.
\end{eqnarray*}
Therefore, by Borel Contelli Lemma, we have that
\begin{eqnarray*}
\liminf\limits_{n\rightarrow\infty}\{\min\limits_{j\in\mathcal{D}}\widehat\omega_j-\max\limits_{j\notin\mathcal{D}}\widehat\omega_j\}\geq\frac{\Delta_\mathcal{E}}{2}>0.
\end{eqnarray*}

\newpage
\section*{References}

\begin{description}

\item Ando, T., $\&$ Li, K. C. (2014). A model-averaging approach for high-dimensional regression. {\sl J. Am. Statist. Assoc.} {\bf 109}, 254-265.

\item Breiman, L. (1995). Better subset regression using the nonnegative garrote. {\sl Technometrics} {\bf 37}, 373-384.

\item Candes, E., $\&$ Tao, T. (2007). The Dantzig selector: statistical estimation when $p$ is much larger than $n$. {\sl Ann. Statist.} 2313-2351.

\item Chang, J., Tang, C. Y., $\&$ Wu, Y. (2013). Marginal empirical likelihood and sure independence feature screening. {\sl Ann. Statist.} {\bf 41}, 2123-2148.

\item Cook, R. D., $\&$ Weisberg, S. (1991). Comment. {\sl J. Am. Statist. Assoc.} {\bf 86}, 328-332.

\item  Cook, R. D., $\&$ Zhang, X. (2014). Fused estimators of the central subspace in sufficient dimension reduction. {\sl J. Am. Statist. Assoc.} {\bf 109}, 815-827.

\item Cui, H., Li, R., $\&$ Zhong, W. (2015). Model-Free Feature Screening for Ultrahigh Dimensional Discriminant Analysis. {\sl J. Am. Statist. Assoc.} {\bf 110}, 630-641.

\item Fan, J., Feng, Y., $\&$ Song, R. (2011). Nonparametric independence screening in sparse ultrahigh-dimensional additive models. {\sl J. Am. Statist. Assoc.} {\bf 106}, 544-557.

\item Fan, J., $\&$ Li, R. (2001). Variable selection via nonconcave penalized likelihood and its oracle properties. {\sl J. Am. Statist. Assoc.} {\bf 96}, 1348-1360.

\item Fan, J., $\&$ Lv, J. (2008). Sure independence screening for ultrahigh dimensional feature space. {\sl J. R.
Statist. Soc. B} {\bf 70}, 849-911.

\item Fan, J., Samworth, R., $\&$ Wu, Y. (2009). Ultrahigh dimensional feature selection: beyond the linear model. {\sl The Journal of Machine Learning Research} {\bf 10}, 2013-2038.

\item Fan, J., $\&$ Song, R. (2010). Sure independence screening in generalized linear models with NP-dimensionality. {\sl Ann. Statist.} {\bf 38}, 3567-3604.

\item Harrion, D., $\&$ Rubinfeld, D. L. (1978). Hedonic prices and the demand for clean air. {\sl  Journal of Environmental Economics and Management} {\bf 5}, 81-102.

\item Hastie, T., $\&$ Tibshirani, R. (1998). Classification by pairwise coupling. {\sl Ann. Statist.} {\bf 26}, 451-471.

\item He, X., Wang, L., $\&$ Hong, H. G. (2013). Quantile-adaptive model-free variable screening for high-dimensional heterogeneous data. {\sl Ann. Statist.} {\bf 41}, 342-369.

\item Hsing, T., $\&$ Carroll, R. J. (1992). An asymptotic theory for sliced inverse regression. {\sl Ann. Statist.} {\bf 20}, 1040-1061.

\item Huang, J., Horowitz, J. L., $\&$ Ma, S. (2008). Asymptotic properties of bridge estimators in sparse high-dimensional regression models. {\sl Ann. Statist.} {\bf 36}, 587-613.

\item Huang, D., Li, R., $\&$ Wang, H. (2014). Feature screening for ultrahigh dimensional categorical data with applications. {\sl Journal of Business $\&$ Economic Statistics} {\bf 32}, 237-244.

\item Li, G., Peng, H., Zhang, J., $\&$ Zhu, L. (2012). Robust rank correlation based screening. {\sl Ann. Statist.} {\bf 40}, 1846-1877.

\item Li, R., Zhong, W., $\&$ Zhu, L. (2012). Feature screening via distance correlation learning.  {\sl J. Am. Statist. Assoc.} {\bf 107}, 1129-1139.

\item Li, K. C. (1991). Sliced inverse regression for dimension reduction.  {\sl J. Am. Statist. Assoc.} {\bf 86}, 316-327.

\item Mai, Q., $\&$ Zou, H. (2015). The Fused Kolmogorov Filter: A Nonparametric Model-Free Screening Method. {\sl Ann. Statist.} {\bf 43}, 1471-1497.

\item Pollard, D. (1984), {\sl Convergence of Stochastic Processes}, Springer-Verlag New York Inc.

\item  Tibshirani, R. (1996). Regression shrinkage and selection via the lasso. {\sl J. R.
Statist. Soc. B} {\bf 58}, 267-288.

\item Yuan, M., $\&$ Lin, Y. (2006). Model selection and estimation in regression with grouped variables. {\sl J. R.
Statist. Soc. B} {\bf 68}, 49-67.

\item  Zhu, L. P., Li, L., Li, R., $\&$ Zhu, L. X. (2011). Model-free feature screening for ultrahigh-dimensional data.  {\sl J. Am. Statist. Assoc.} {\bf 106}, 1464-1475.

\item Zhu, L., $\&$ Ng, K. W. (1995). Asymptotics of sliced inverse regression. {\sl Statist. Sinica} {\bf 5}, 727-736.

\item Zou, H. (2006). The adaptive lasso and its oracle properties. {\sl J. Am. Statist. Assoc.} {\bf 101}, 1418-1429.

\end{description}

\newpage
{\renewcommand{\arraystretch}{0.6} \tabcolsep0.04 in
\begin{center}
{Table 1(a). The median of the minimum model sizes (MMS) for 300 replications in the simulation study}\\[2mm]
 \begin{tabular}{ccccccccccccc}\hline \hline
Expert. & $N^\#$ &FMV&SIS&DCS&RCS&NIS&FKS&QAS(0.5)&QAS(0.75)&SIRS&ELS \\ \hline
 (1a)&8&8&8& 8 & 8&8&8&8&8&8&8\\
 (1b)&8&8 &11 & 8 &8 &497 &8 &8 &8&8&15 \\
 (1c)&2&2&2&2 &2&2 &2&2&2&2&2 \\
 (1d)&2&2&601&30&2&718 &2&10&709&2&604 \\
 (2a)&2&2&1422&287&2&2&2&45.5&80.5&1120&1788 \\
 (2b)&2&2&398.5&301& 2 & 1138&2 &8&39.5&2&2057 \\
 (2c)&8&8&336& 61.5 & 8&890.5&8&402.5&307&8&2281 \\
 3&3&3&300.5& 231 & 3&398.5 &3&41&72.5&389.5&2611 \\
 4&3&3&1865&15& 3 & 2157.5 &3&1665.5&2023&3&1462 \\
 5&8&11&1891&24 & 1122.5&1920 &14&303.5&27.5&19&519 \\
 6& 2&2 &25.5 & 41 & &  &12.5 & &&&2077  \\ 
7&4&4&11$^\flat$ &  189$^\natural$&\\ \hline
\end{tabular} \\ \vspace{3mm}
\footnotesize{Note: $N^\#$ denotes the true number of the active predictors, 
$11^\flat$ denotes the MMS value of the ``naive'' QAS(0.3) screening procedure,
and $189^\natural$ denotes the MMS value of the Cox(SIS) screening procedure.}
\end{center}
}

{\renewcommand{\arraystretch}{0.6} \tabcolsep0.04 in
\begin{center}
{Table 1(b). Standard errors of MMS values for 300 replications in the simulation study}\\[2mm]
 \begin{tabular}{ccccccccccccc}\hline \hline
Expert. &$N^\#$ &FMV&SIS&DCS&RCS&NIS&FKS&QAS(0.5)&QAS(0.75)&SIRS&ELS \\ \hline
 (1a)&8&0&0& 0 & 0&0 &0&0&0&0&0\\
 (1b)&8&0.32&153.1& 0.38& 0.36& 191.7&0.32 & 0.4&58&0.33&3.2 \\
 (1c)&2&0&0& 0 & 0&0 &0&1.2&3.5&0&0.6 \\
 (1d)&2&0.1&269& 112.2& 0.2&218.9 &0.12&34.7&298.6&16.3&69.9 \\
(2a)&2&0&142.3&88.7&0.4&1.4&0&7.8&38.4&223.1&94.3 \\
(2b)&2&0.2&161.5&83.2& 0& 121.3&0.23 &1.4&10.2&0.5&102.2 \\
(2c)&8&0.4&766.2&253.7 & 0.2&322.2&0.5&10.2&33.2&0.4&87.3 \\
 3&3&2.5&44.5&52.2 &3& 123.7&2.3 &3.6&5.2&323.1&88.1 \\
 4&3& 5.5&885.2&54.2&4.2&735 &6.9 &962&718&821&91.3 \\
 5&8&3.4&521.2& 7.3 & 179.1&122.9 &3.1 & 321.1&4.2&12.1&104.4 \\
 6&2&0 &3.2 & 17.9 & & & 4.2& & &&132.2 \\
 7&4&20.9&24.1$^\flat$& 331.2$^\natural$&\\ \hline
\end{tabular} \\ \vspace{3mm}
\end{center}
}

{\renewcommand{\arraystretch}{0.6} \tabcolsep0.04 in
\begin{center}
{Table 2. Comparison of the screening methods on the Boston Housing data. We report the number of
true predictors that are preserved after the screening step. The numbers are averaged
over 100 replicates. Standard errors are below}\\[2mm]
 \begin{tabular}{ccccccccccccc}\hline \hline

&  &FMV&SIS&SIRS&FKS&NIS&DCS \\ \hline
&True predictors&89.4&83.2&73.2&86.6&65.8&77.9\\
&standard errors&0.24&0.76&2.42& 0.45&1.91&0.91 \\
\hline
\end{tabular} \\ \vspace{3mm}
\end{center}
}
\begin{figure}
  \centering
  \includegraphics[width=16cm,height=15cm]{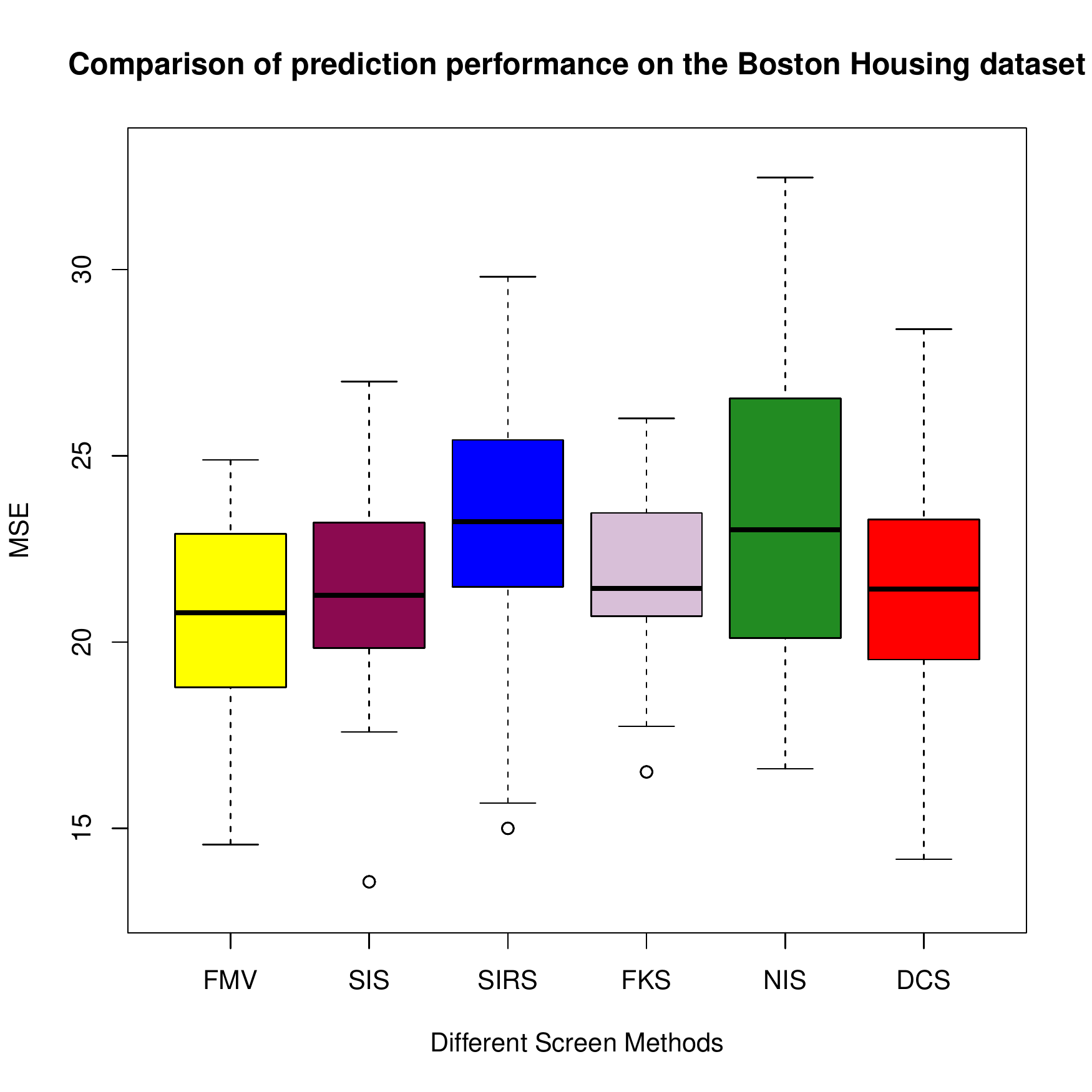}\\
  \caption{ Boxplots of the prediction performance on the Boston Housing data
  }\label{tex}
\end{figure}

\end{document}